\documentclass[journal]{IEEEtran}
\usepackage{amsmath,amssymb}
\usepackage{subfigure}
\usepackage{graphicx,graphics,color,psfrag}
\usepackage{cite,balance}
\usepackage{caption}
\allowdisplaybreaks
\usepackage{algorithm}
\usepackage{algorithmic}
\usepackage{accents}
\usepackage{amsthm}
\usepackage{bm}
\usepackage{url}
\usepackage[english]{babel}
\usepackage{multirow}
\usepackage{enumerate}
\usepackage{cases}
\usepackage{stfloats}
\usepackage{dsfont}
\usepackage{color,soul}
\usepackage{amsfonts}
\usepackage{cite,graphicx,amsmath,amssymb}
\usepackage{subfigure}
\usepackage{fancyhdr}
\usepackage{hhline}
\usepackage{graphicx,graphics}
\usepackage{array,color}
\usepackage{mathtools}
\usepackage{amsmath}

\newtheorem{definition}{\emph{\underline{Definition}}}

\newtheorem{observation}{\emph{\underline{Observation}}}

\newtheorem{remark}{\bf \emph{\underline{Remark}}}

\def\({\left(}
\def\){\right)}

\def\b0{{\mathbf{0}}}

\newcommand{\nn}{\nonumber}

\begin{document}
\captionsetup[figure]{name={Fig.}}

\title{\huge 
Fast Near-Field Beam Training for Extremely \\ Large-Scale Array} 
\author{Yunpu Zhang, Xun Wu, and  
        Changsheng You, \IEEEmembership{Member, IEEE}
\thanks{The authors are with the Department of Electronic and Electrical Engineering, Southern University of Science and Technology (SUSTech), Shenzhen 518055, China (e-mail: zhangyp2022@mail.sustech.edu.cn; 11812112@mail.sustech.edu.cn; youcs@sustech.edu.cn).}}
\maketitle

\begin{abstract} 
In this letter, we study efficient near-field beam training design for the extremely large-scale array (XL-array) communication systems. Compared with the conventional far-field beam training method that searches for the best beam direction only, the near-field beam training is more challenging since it requires a beam search over both the angular and distance domains due to the spherical wavefront propagation model. To reduce the near-field beam-training overhead based on the two-dimensional exhaustive search, we propose in this 
letter a new \emph{two-phase} beam training method 
that decomposes the two-dimensional search into two sequential phases. Specifically, in the first phase, the candidate angles of the user is determined by a new method based on the conventional far-field codebook and angle-domain beam sweeping. Then, a customized polar-domain codebook is employed in the second phase to find the best effective distance of the user given the shortlisted candidate angles.
Numerical results show that our proposed two-phase beam training method significantly reduces the training overhead of the exhaustive search and yet achieves comparable beamforming performance for data transmission.
\end{abstract}
\vspace{-0.5cm}
\section{Introduction}
\vspace{-0.1cm}
Extremely large-scale array/surface (XL-array/surface) with active or passive elements has emerged as a promising technology to significantly enhance the spectral efficiency and spatial resolution in future sixth-generation (6G) wireless network \cite{cui2022near1,9326394,9617121,zheng2022simultaneous}. 
Compared with the massive multiple-input-multiple-output (MIMO) for fifth-generation (5G), XL-array for 6G introduces several new channel characterizations. Specifically, with the significant increase of the antenna number and carrier frequency, the users are more likely to locate in the near-field region with spherical wavefront propagation, due to the smaller user/link distance than the enlarged \emph{Rayleigh distance}. As such, the conventional transceiver designs based on the far-field channel model with the planar wavefront assumption may not be applicable, hence calling for new design approaches dedicated to the near-field communication \cite{luo2022beam,8949454,wei2021codebook,cui2022channel,cui2022near}.

In particular, to reap the prominent  beamforming gain brought by the XL-array, it is indispensable for the XL-array to perform the near-field beam training for establishing high signal-to-noise ratio (SNR) initial links before efficient channel estimation and data transmission \cite{wei2021codebook,you2020fast}. In contrast to the conventional far-field beam training methods that aim to find the best beam direction based on a predefined angular-domain beam codebook, the near-field beam training requires new codebook designs and efficient search for a proper beam to match the specific channel angle and distance according to the spherical wavefront model. As such, directly applying the conventional far-field beam training methods\cite{7390101,you2020fast} to the near-field communication systems will result in considerable performance degradation, especially when the user is very close to the XL-array. 

To address the above issues, several new beam training methods dedicated to the near-field communication have been recently proposed in the literature \cite{cui2022channel,wei2021codebook,cui2022near}. Specifically, the authors in \cite{wei2021codebook} proposed a hierarchical near-field codebook and the corresponding hierarchical near-field beam training scheme, where different levels of sub-codebooks are searched in turn with
reduced codebook size. To further reduce the size of the near-field codebook, a novel polar-domain codebook dedicated to the near field was proposed in \cite{cui2022channel}, wherein the angular domain is uniformly sampled while the distance domain is sampled non-uniformly to reduce the codebook size. In addition, by exploiting the near-field beam squint effect, a fast wideband beam training scheme was proposed in \cite{cui2022near}. Specifically, by operating beams at different frequencies to be focused on different locations, the optimal beamforming vector is rapidly searched, thus greatly reducing the training overhead. However, the existing near-field training methods for the narrow-band require a two-dimensional exhaustive search for all possible beam directions and distances, thus leading to prohibitively high training overhead.

Motivated by the above, we consider an XL-array communication system and propose a new \emph{two-phase} fast beam training method to significantly reduce the near-field beam training overhead based on the exhaustive search. To this end, we first show that the actual user spatial direction approximately locates in the middle of a newly defined dominant-angle region. Based on this key observation, 
we propose to decompose the two-dimensional search into two sequential phases. Specifically, the first phase determines the candidate angles of the user based on the far-field codebook and the angle-domain beam sweeping. Then, the second phase employs a customized polar-domain codebook to find the best effective distance of the user given the shortlisted candidate angles instead of the full angular domain as in the exhaustive search, thus effectively reducing the training overhead.
 Finally, numerical results show that the proposed fast beam training scheme greatly reduces the training overhead of the exhaustive search, yet without compromising much the XL-array beamforming performance for data transmission.
\vspace{-0.9cm}
\section{System model}
\begin{figure}
    \centering
    \includegraphics[width=6cm]{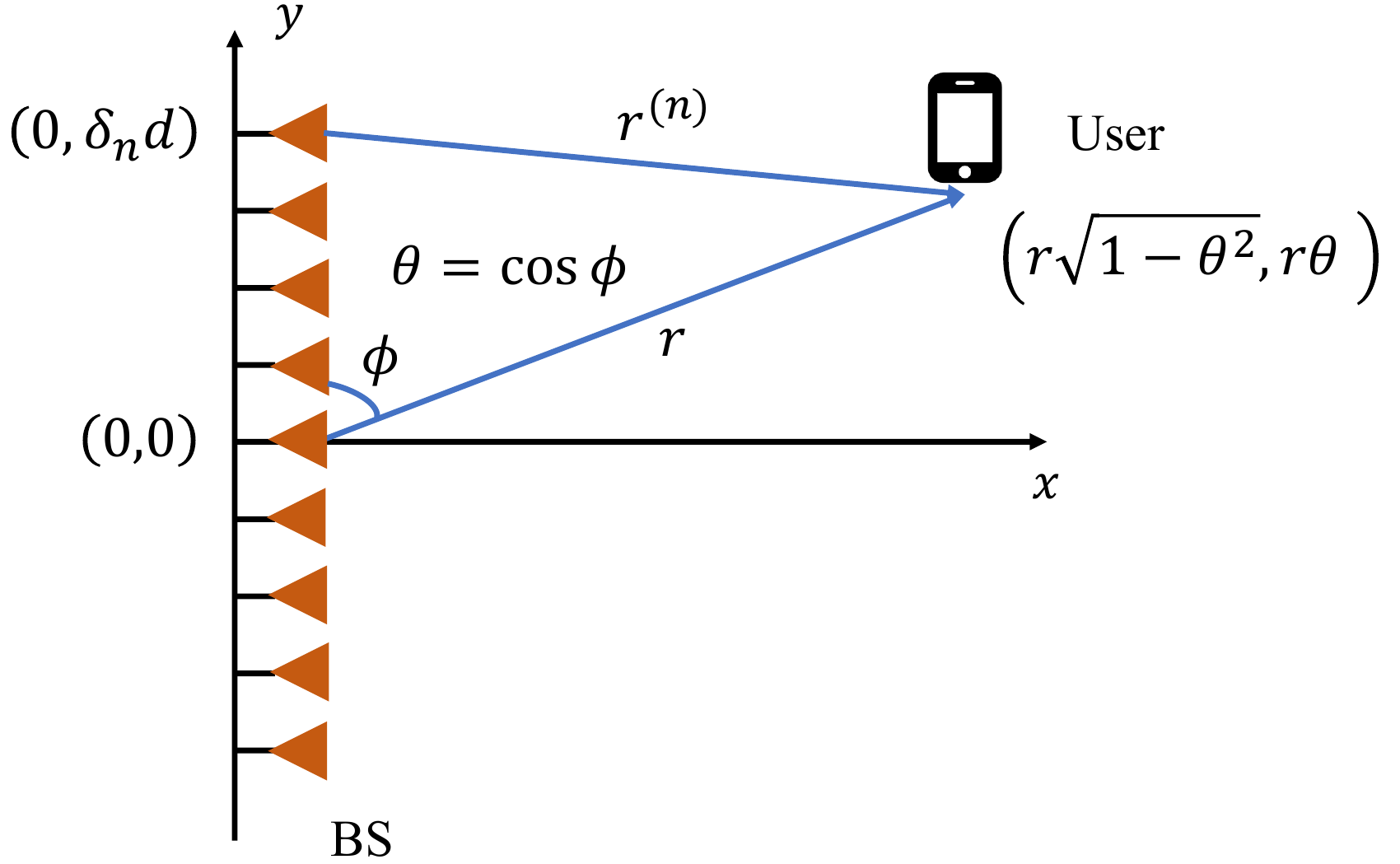}
    \caption{A narrow-band XL-array communication system.}
    \label{fig:sytem_model}
    \vspace{-16pt}
\end{figure}

\vspace{-0.3cm}
As shown in Fig.~\ref{fig:sytem_model}, we consider the downlink beam training for a narrow-band XL-array communication system, where a base station (BS) equipped with $N$-antenna uniform linear array (ULA) communicates with a single-antenna user.\footnote{The proposed two-phase beam training scheme can be easily extended to the multi-user scenario by performing the angle estimation for all users at the same time and sequentially estimating the effective distance for each user.}

\underline{\bf Near-field channel model:}
Generally speaking, the electromagnetic radiation field can be divided into the far-field and near-field regions, leading to different channel characterizations \cite{9738442}. Specifically, the user is assumed to locate in the far-field or near-field region when the BS-user distance is larger or smaller than the well-known Rayleigh distance, denoted by $Z=\frac{2D^2}{\lambda}$ with $D$ and $\lambda$ denoting the antenna array aperture and the carrier wavelength, respectively. In particular, for XL-array communication systems in high-frequency bands (e.g., mmWave/Terahertz), say $D=0.4$ m and $f = 100$ GHz, the Rayleigh distance is about 107 m and thus the user usually locates in the near field, whose channel is modeled as follows.

Without loss of generality, the $N$-antenna BS is placed at the $y$-axis with the coordinate of the $n$-th antenna given by ($0, \delta_{n}d$), where $\delta_{n}=\frac{2n-N-1}{2}$ with $n=1,2,\cdots,N$, and $d=\frac{\lambda}{2}$ is the antenna spacing. Then, based on the spherical wavefront propagation model, the near-field line-of-sight (LoS) channel from the BS to the user, denoted by $\mathbf{h}^H_{\rm near}$, can be modeled as \cite{cui2022channel}\footnote{For the general multi-path channel model, new near-field beam training scheme need to be developed catered to the complicated beam pattern, which is more challenging and left for our future work.}
\begin{equation}\label{Eq:nf-model}
    \mathbf{h}^H_{\rm near}=\sqrt{N}h \mathbf{b}^{H}(\theta,r),\vspace{-3pt}
\end{equation}
where $h=\frac{\sqrt{\beta}}{r} e^{-\frac{\jmath 2 \pi r}{\lambda}}$ is the complex-valued channel gain with $\beta$ and $r$ denoting the reference channel gain at a distance of $1$ m and the distance between the BS center and the user, respectively. $\mathbf{b}^{H}(\theta,r)$ denotes the near-field steering vector, which is given by
\begin{equation}\label{near_steering}
    \mathbf{b}^{H}\left(\theta, r\right)=\frac{1}{\sqrt{N}}\left[e^{-\jmath 2 \pi(r^{(0)}-r)/\lambda}, \cdots, e^{-\jmath 2 \pi(r^{(N-1)}-r)/\lambda}\right],
\end{equation}
where $\theta$ denotes the spatial angle at the BS, given by $\theta=2d\cos(\phi)/ \lambda$ with $\phi$ denoting the physical angle-of-departure (AoD) from the BS center to the user; $r^{(n)}=\sqrt{r^2+\delta_{n}^2d^2-2r\theta \delta_{n}d}$ represents the distance between the $n$-th antenna at the BS (i.e., ($0, \delta_{n}d$)) and the user.

\underline{\bf Signal model:} Based on \eqref{Eq:nf-model}, the received signal at the user is given by
\begin{align}\label{Eq:general_Sig}
    y_{\rm near} = \mathbf{h}^H_{\rm near}\mathbf{v}x+z_{0}
    =\sqrt{N}h \mathbf{b}^{H}(\theta,r)\mathbf{v}x+z_{0},
    \vspace{-3pt}
\end{align}
where $x\in \mathbb{C}$ denotes the transmitted symbol by the BS with power $P$, $\mathbf{v}\in \mathbb{C}^{N \times 1}$ represents the transmit beamforming vector at the BS based on the power-efficient analog phase shifters \cite{cui2022near1}, and $z_{0}$ is the received additive white Gaussian noise (AWGN) at the user with power $\sigma^2$.

For the near-field beam training, it can be easily obtained that the optimal BS beamforming vector is $\mathbf{v}_{ \rm opt}=\mathbf{b}(\theta,r)$. This means that the optimal beam-training codeword should align both the spatial angle $\theta$ and BS-user distance $r$, thus  significantly differs from the conventional far-field beam training method, which will be detailed in the next section.
\vspace{-6pt}
\section{Benchmark schemes}
\vspace{-3pt}
In this section, we introduce two benchmark beam-training schemes and point out their main issues.
\vspace{-8pt}
\subsection{Angle-Domain Far-field Beam Training Based on Exhaustive Search}
First, consider the conventional far-field beam training method and its issues when directly applied in the near-field beam training. Based on the planar wavefront assumption, the far-field channel model $\mathbf{h}_{\rm far}^H$ can be characterized as 
\begin{equation}\label{Eq:far}
    \mathbf{h}^H_{\rm far}=\sqrt{N} h\mathbf{a}^{H}(\theta),
    \vspace{-3pt}
\end{equation}
where $\mathbf{a}^{H}(\theta)$
denotes the far-field steering vector, given by
\begin{equation}
    \mathbf{a}^{H}(\theta)\triangleq \frac{1}{\sqrt{N}}\left[1, e^{-\jmath \pi \theta},\cdots, e^{-\jmath \pi (N-1)\theta}\right].
    \vspace{-3pt}
\end{equation}
Similar to \eqref{Eq:general_Sig}, the received signal at the user can be characterized as
\begin{equation}
     y_{\rm far} = \mathbf{h}^H_{\rm far}\mathbf{v}x+z_{0}
     =\sqrt{N} h\mathbf{a}^H(\theta)\mathbf{v}x+z_{0}.
     \vspace{-3pt}
\end{equation}
For the classical far-field beam training, the entire spatial domain $[-1, 1]$ is divided into $N$ equal-size sectors, represented by their central directions specified by $\theta_{n}=\frac{2n-N-1}{N}, n \in \mathcal{N}\triangleq \{1,2,\cdots, N\}$. As such, the far-field codebook consisting of $N$ codewords can be constructed as $\mathbf{W}=\{\mathbf{w}_{1},\mathbf{w}_{2},\cdots, \mathbf{w}_{N}\}$, each codeword $\mathbf{w}_{n}$ corresponding to a transmit beam pointing to the central direction $\theta_{n}$, i.e., $\mathbf{w}_{n}=\mathbf{a}(\theta_{n})$ \cite{you2020fast}. Then, the BS sequentially selects 
its $N$ codewords as its transmit beamforming vector over the training symbols. Last, the user measures its received signal power over time and feeds back the best codeword index that leads to the maximum received signal power.

However, the conventional far-field beam training method cannot be applied to the considered near-field communication. This is because the far-field channel model defined in \eqref{Eq:far} mainly depends on the spatial angle $\theta$, while the near-field steering vector in \eqref{near_steering} is jointly determined by both the distance and spatial angle. Thus, straightforwardly applying the far-field beam training method to the near-field communication scenario will result in significant rate performance loss due to the channel-model mismatch (see Section~\ref{SimuR}).

\vspace{-6pt}
\subsection{Polar-Domain Near-field Beam Training Based on Exhaustive Search}
For the near-field beam training, we consider the typical polar-domain near-field codebook proposed in \cite{cui2022channel}, 
which is given by $\mathbf{F}=\{\mathbf{F}_1,\mathbf{F}_2,\cdots,\mathbf{F}_N\}$ with $\mathbf{F}_n=\{\mathbf{f}_{n,0},\mathbf{f}_{n,1},\cdots,\mathbf{f}_{n,S_{n}-1}\}, \forall n\in \mathcal{N}$,
wherein each codeword $\mathbf{f}_{n,s_{n}}, s_{n} \in \mathcal{S}_{n} \triangleq\{0,1,\cdots,S_{n}-1\}$ corresponds to a beam pointing to a specific \emph{direction-and-distance} pair (i.e., $(\theta_{n},r_{n,s_{n}})$), which is given by $\mathbf{f}_{n,s_{n}}=\mathbf{b}(\theta_{n},r_{n,s_{n}})$. Specifically, the spatial domain is equally divided into $N$ sectors represented by its central angle; while in the distance domain, for each angle $\theta_{n}$, the distance is divided into $S_{n}$ (which may not be the same for different $\theta_{n}$) non-uniform regions where the distance sampling interval gets larger as the distance increases. This is expected since the distance effect on the beamforming design diminishes with the distance and finally disappears in the far-field region (i.e., only determined by the spatial angle). Based on the proposed polar-domain codebook \cite{cui2022channel},
the exhaustive-search based beam training method can be applied to find the best beam codeword for the user.
However, this exhaustive-search method requires a total of $\sum_{n=1}^{N}S_{n}$ training symbols, which is prohibitively high when $N$ is large and hence results in insufficient/less time for data transmission.

\vspace{-0.2cm}
\section{Proposed two-phase near-field beam training Method}
\vspace{-0.1cm}
To reduce the huge training overhead of the near-field beam training method based on the exhaustive search, in this section, we propose a new two-phase beam training method and show its appealing advantages as compared to the benchmark schemes.

First, we make the following key definitions.
\begin{definition}
\emph{Define $\mathbf{A}(\mathbf{u}^H,\mathbf{w})$ as the normalized beam gain of the beamforming vector $\mathbf{w}$ along the channel steering vector $\mathbf{u}^H$, which is given by
\begin{equation}
    \mathbf{A}(\mathbf{u}^H,\mathbf{w})=|\mathbf{u}^H\mathbf{w}|.
    \vspace{-3pt}
\end{equation}}
\end{definition}
\begin{definition}\label{De:region}
\emph{Define $\mathbf{w}(\Omega)$ as the far-field beamforming vector function for the spatial angle $\Omega\in [-1,1]$. Then, let  $\mathcal{C}(\mathbf{b}^H(\theta,r),\mathbf{w})$ denote the \emph{dominant-angle region} for $\mathbf{b}^H(\theta,r)$ that characterizes the spatial angle region where using the far-field beamforming vector $\mathbf{w}$ can lead to sufficiently high beam power for $\mathbf{b}^H(\theta,r)$. Mathematically, $\mathcal{C}(\mathbf{b}^H(\theta,r),\mathbf{w})$ is given by
\begin{align}
    \mathcal{C}(\mathbf{b}&^H(\theta,r),\mathbf{w})=\nn\\
    &\{\Omega| \mathbf{A}(\mathbf{b}^H(\theta,r),\mathbf{w}(\Omega))>\rho \max_{\mathbf{w}} \mathbf{A}(\mathbf{b}^H(\theta,r),\mathbf{w})\},
\end{align}
where $\rho \in (0,1)$ is the beam gain threshold for $\mathbf{b}^H(\theta,r)$. In particular, $\rho$ can be set as $1/\sqrt{2}$, corresponding to the $3$ dB dominant-angle region \cite{7390101}.}
\end{definition}

\begin{figure}
    \centering
    \includegraphics[width=6cm]{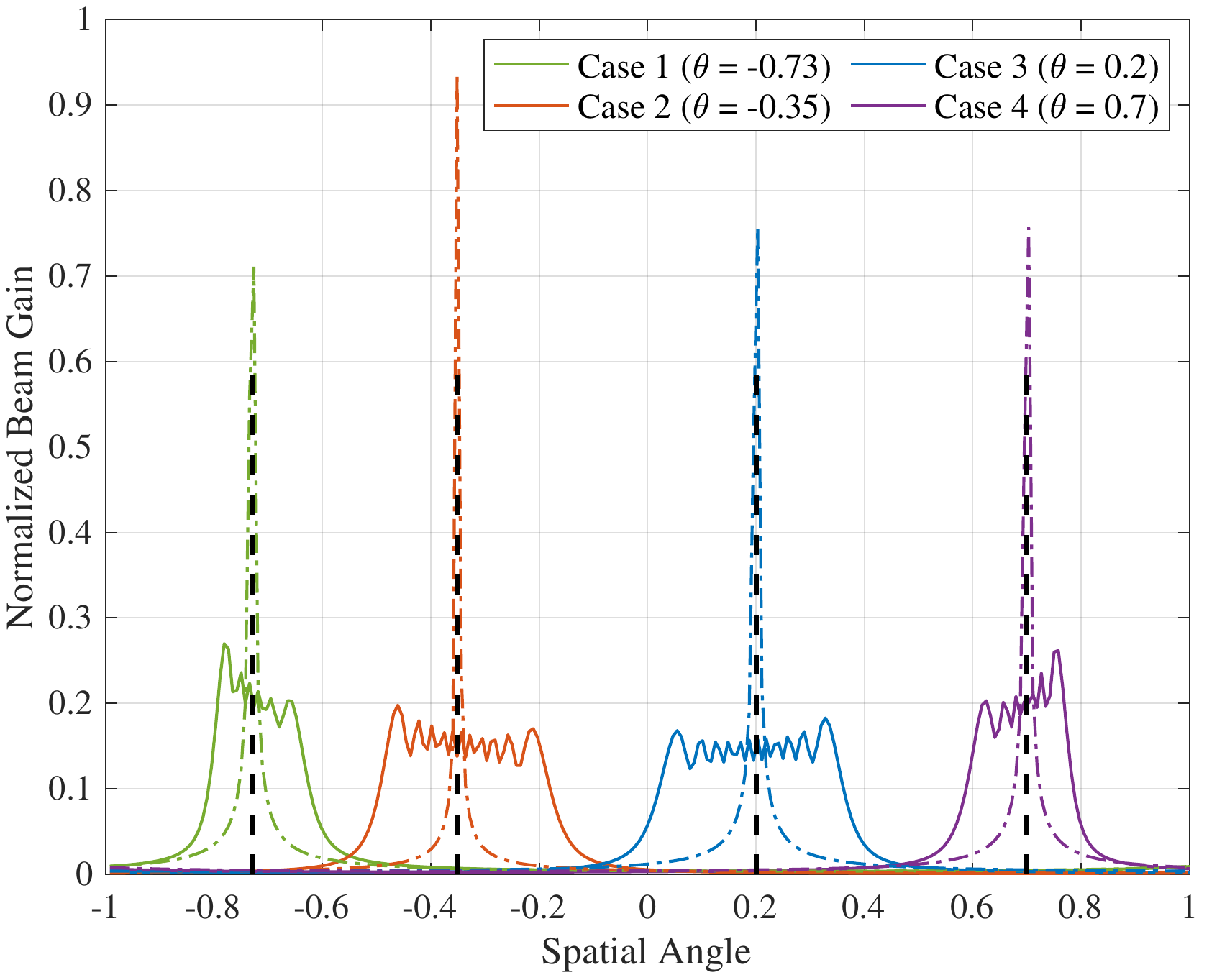}
    \caption{Normalized beam gains of $\mathbf{w}(\Omega)$ along the far- (marked by dotted lines) versus near-field (marked by solid lines) steering vectors, i.e., $\mathbf{a}^H(\cdot)$ and $\mathbf{b}^H(\cdot)$, under four angles (marked by black dashed lines), where the BS antenna number is 256 and the carrier frequency is 100 GHz. The BS-user distances of the near-field and far-field are set as $1$ m and $100$ m, respectively.}
    \label{fig:near_far_field}
    \vspace{-16pt}
\end{figure}

Then, we show in Fig.~\ref{fig:near_far_field} the normalized beam gains of the far-field beamforming vector $\mathbf{w}$ versus the spatial angle under both the far- and near-field channel models, which are given by $\mathbf{A}(\mathbf{h}^H_{\rm far},\mathbf{w})=|\mathbf{a}^{H}(\theta)\mathbf{w}|$ and $\mathbf{A}(\mathbf{h}^H_{\rm near},\mathbf{w})=|\mathbf{b}^{H}\left(\theta, r\right)\mathbf{w}|$, respectively. One key observation is highlighted as follows.
\begin{observation}[angle-in-the-middle]\label{obs}
\emph{In Fig.~\ref{fig:near_far_field}, the true spatial angle $\theta$ approximately locates in the middle of the dominant-angle region based on the far-field beamforming vector $\mathbf{w}$. Mathematically, we have $ \theta \approx {\rm Med}(\mathcal{C}(\mathbf{b}^H(\theta,r),\mathbf{w}))$. }
\end{observation}

\noindent\textbf{Observation~\ref{obs}} indicates that the true spatial angle for the near-field user can be approximated located by using the conventional far-field beam training method. This thus motivates the proposed design of a new two-phase near-field beam training method. The key idea is to firstly estimate the candidate spatial angle based on the far-field codebook, following by the effective distance estimation in the second phase based on 
a customized polar-domain codebook. The detailed procedures for the two phases are elaborated as follows.
\begin{itemize}
    \item \textbf{First phase (Candidate angle estimation):} 
    Motivated by \textbf{Observation~\ref{obs}}, we aim to estimate the spatial angle of the user in the first phase based on the far-field beam training method. Specifically, an angle-domain codebook $\mathbf{W}$ is constructed with each codeword pointing to one spatial direction.
    
    1) \emph{Angle-domain beam sweeping:} The BS sequentially sends $N$ training symbols, while it dynamically tunes its beam direction (specified by the codeword) according to the predefined angle-domain codebook $\mathbf{W}=\{\mathbf{w}_{1},\mathbf{w}_{2},\cdots, \mathbf{w}_{N}\}$.
     For each codeword $\mathbf{w}_{n}$, the received signal at the user is given by
    \begin{equation}
        y(\mathbf{w}_{n})=\sqrt{N}h \mathbf{b}^H(\theta,r)\mathbf{a}(\theta_{n})x+z_{0}.
    \end{equation}
    2) \emph{Candidate angle estimation:}
    After the beam sweeping, based on \textbf{Definition~\ref{De:region}}, the index set of the codewords for which the user receives significantly high power (i.e., dominant-angle region), denoted by $\mathbf{\Psi}$, is given by $\mathbf{\Psi}=\{ n| \;|y(\mathbf{w}_n)|^2> \rho^2 \max \limits_{\mathbf{w}} |y(\mathbf{w})|^2\}$.
 
    Note that the median angle of the obtained dominant-angle region may not be accurate enough due to the power fluctuation, the existence of received noise and the approximation in \textbf{Observation~\ref{obs}}. To address this issue, we thus propose a new \emph{middle-$K$ angle selection scheme} that selects $K$ candidate spatial angles in the middle of the \emph{quantized dominant-angle region} rather than selecting one spatial angle only. Specifically, let $\Xi$ denote the \emph{candidate angle index set}, which is given by
    $\Xi=\{\lfloor{\rm Med}(\mathbf{\Psi})\rfloor-\lfloor\frac{K-1}{2}\rfloor,\cdots,\lfloor{\rm Med}(\mathbf{\Psi})\rfloor,\cdots,\lfloor{\rm Med}(\mathbf{\Psi})\rfloor+\lceil\frac{K-1}{2}\rceil\}$ with $K$ denoting the total number of candidate spatial angles.
    \item \textbf{Second phase (effective distance estimation)}: Based on the obtained candidate-angle index set $\Xi$ in the first phase, a customized polar-domain beam training method is proposed for the second phase to estimate the \emph{effective user distance} based on the non-uniform distance sampling method \cite{cui2022channel}. Note that the effective user distance is defined as the best sampled distance for the user in the polar-domain codebook, which may not be the true BS-user distance due to the limited number of sampling. Specifically, the polar-domain codebook $\mathbf{F}$ is adopted which consists of $\sum_{n=1}^{N}S_{n}$ sets of codewords, each corresponding to one specific angle-and-distance pair.
    
    1) \emph{Distance-domain beam sweeping:} The BS sends $K$ sets of pilot symbols to the user. For each candidate angle $\theta_{k}, k \in \Xi$, the sampled distances are given by
      \begin{equation}\label{distance_rule}
        r_{k,s_{k}}=\frac{1}{s_{k}}Z_{\Delta}(1-\theta_{k}^{2}), \; s_{k}=0,1,2,3,\cdots,
        \vspace{-4pt}
    \end{equation}
     where $Z_{\Delta}=\frac{D^{2}}{2 \alpha_{\Delta}^{2} \lambda}$ is the threshold distance defined to limit the column coherence between two near-field steering vectors \cite{cui2022channel}.
     The corresponding non-uniform distance index set, denoted by $\mathcal{R}_{k}$, is given by
    \begin{equation}
        \mathcal{R}_{k}\triangleq\{0,\cdots,s_{k},\cdots,S_{k}-1\}.
        \vspace{-3pt}
    \end{equation}
    Then, given the polar-domain codeword $\mathbf{f}_{k,s_{k}}$, the received signal at the user is given by
    \begin{equation}
        y(\mathbf{f}_{k,s_{k}})=\sqrt{N}h \mathbf{b}^H(\theta,r)\mathbf{b}(\theta_{k},r_{k,s_{k}})x+z_{0}.\vspace{-4pt}
    \end{equation}
    
     2) \emph{Beam determination:} After the distance-domain beam sweeping in the second phase, the user determines its best polar-domain beam codeword, denoted by $\hat{\mathbf{f}}_{k^*,s^*_{k^*}}$ based on its received signal powers/SNRs. 
     Mathematically, the best polar-domain codeword index is given by
       \begin{equation}
        (k^*,s^*_{k^*})=\arg \max_{k \in \Xi, s_{k}\in \mathcal{R}_{k}} | y(\mathbf{f}_{k,s_{k}})|^2.
        \vspace{-4pt}
    \end{equation}
    The best beam direction-and-distance is thus given by $(\theta_{k^*},r_{k^*,s^*_{k^*}})$. 
\end{itemize}
The detailed steps for the proposed two-phase beam training method are summarized in Algorithm~\ref{Alg:Two-phase}.
\begin{algorithm}[t]
\caption{Proposed Two-phase Beam Training Method}
\label{Alg:Two-phase}
\begin{algorithmic}[1]
\STATE \textbf{Phase 1: Candidate angle estimation}
\STATE Use the far-field codebook $\mathbf{W}$ for the angle-domain beam sweeping and obtain the dominant-angle region $\mathbf{\Psi}$.
\STATE Obtain $\Xi$ according to the proposed middle-$K$ angle selection scheme.
\STATE \textbf{Phase 2: Effective distance estimation:}
\FOR{$k \in \Xi$}
\FOR{$s_{k} \in \mathcal{R}_{k}$}
\STATE $ y(\mathbf{f}_{k,s_{k}})=\sqrt{N}h \mathbf{b}^H(\theta,r)\mathbf{b}(\theta_{k},r_{k,s_{k}})x+z_{0}$
\ENDFOR
\ENDFOR
\STATE \textbf{Output:} The best polar-domain codeword $\hat{\mathbf{f}}_{k^*,s^*_{k^*}}$.
\end{algorithmic}
\end{algorithm}
\vspace{-0.2cm}
\begin{remark}[Accuracy of effective distance estimation]\label{Re:dist}
\emph{Based on the non-uniform distance sampling method in \eqref{distance_rule}, the distance domain is more frequently sampled in the short-distance region and less sampled otherwise. This indicates that the effective distance estimation is more accurate when the user is nearer the BS, while it may incur a large error in the long-distance region. This error is expected since the effective distance is resolved from the best beam-training vector, while the distance has the negligible effect in the far-field region. Fortunately, it is worth noting that the possible inaccurate distance estimation due to the non-uniform distance sampling will not affect too much the beamforming performance of the proposed beam training method, which is the ultimate goal of the near-field beam training and will be numerically validated in Section \ref{SimuR}.}
\end{remark}
\vspace{-0.2cm}
\begin{remark}[Universal in both
near- and far-field communications]\label{Re:both}
\emph{It is worth mentioning that the proposed two-phase beam training method can be easily extended to a universal beam training method applicable to both the near- and far-field communications. Specifically, after the first phase (candidate angle estimation), the user can decide whether the received signal power is dispersed in a large angle domain. If the user receives strong signal power only in a small number of training symbols (e.g., only one), it indicates that the user is likely to locate in the far-field region and thus it does not need to execute the second phase. Otherwise, it goes to the second phase for the near-field effective distance estimation.}
\end{remark}
\vspace{-0.2cm}
\begin{remark}[Training overhead]
{\emph{The total number of training symbols of our proposed two-phase beam training method, denoted by $T^{\rm (2P)}$, is given by
\vspace{-0.2cm}
\begin{equation}
    T^{\rm (2P)}=T^{\rm (ae)}+T^{\rm (de)}=N+\sum_{k=1}^{K}S_{k},
    \vspace{-8pt}
\end{equation}
where $T^{\rm (ae)}=N$ denotes the training overhead of the first phase (candidate angle estimation) and $T^{\rm (de)}=\sum_{k=1}^{K}S_{k}$ denotes the training overhead of the second phase (effective distance estimation). 
It can be easily verified that $T^{\rm (2P)}$ is much smaller than the training overhead of the exhaustive-search based near-field beam training method \cite{cui2022channel}, i.e., $T^{\rm (ex)}=\sum_{n=1}^{N}S_{n}$ .
For instance, consider the case where $N=256$, $f=100$ GHz, $S_{n}=S_{k}=6, \forall n\in \mathcal{N}, k\in \Xi$ \cite{cui2022channel}, and $K=3$. Then, we have $T^{\rm (2P)}=274$, which is much smaller than $T^{\rm (ex)}=1536$. Last, it is worth mentioning that there exists a fundamental tradeoff between the beam training performance versus the number of candidate angles, $K$. Specifically, when $K$ is larger, it incurs a larger training overhead, while leading to a higher beamforming gain due to the more accurate angle estimation.}}
\end{remark}
\vspace{-0.6cm}
\section{Numerical results}\label{SimuR}
In this section, we provide numerical results to validate the effectiveness of our proposed two-phase beam training method. Specifically, we consider a XL-MIMO communication system with $N=256$, $\alpha_{\Delta}=1.2$\cite{cui2022channel}, $f=100$ GHz, and $\beta=(\lambda/4\pi)^2=-72$ dB. We define the (reference) SNR of the XL-MIMO system as ${\rm SNR}=\frac{P\beta N}{r^2\sigma^2}$, where the noise power is set as $\sigma^2=-70$ dBm. To characterize the beam determination accuracy, we first define $(\bar{n},\bar{s}_{\bar{n}})$ as the index of the best polar-domain codeword, which is given by $(\bar{n},\bar{s}_{\bar{n}})=\arg \max \limits_{\mathbf{f}\in \mathbf{F}} |\mathbf{h}^H_{\rm near}\mathbf{f}|^2$. As such, we define the (beam-training) success rate as $P_{\rm suc}=\frac{\sum^{M}_{m=1}\mathbb{I}(
\bar{n}=k^*)\cdot \mathbb{I}(\bar{s}_{\bar{n}}=s^*_{k^*})}{M}$, where $\mathbb{I}(\cdot)$ stands for the indicator function and $M$ denotes the number of beam training realizations which is set as $M=1000$. Moreover, the achievable rate is give by $R=\log_{2}(1+\frac{P\beta N|\mathbf{b}^H(\theta,r)\mathbf{v}|^2}{r^2\sigma^2})$.
\begin{figure}[t]
\centering
\subfigure[{Success rate versus SNR}]{\includegraphics[width=4.3cm]{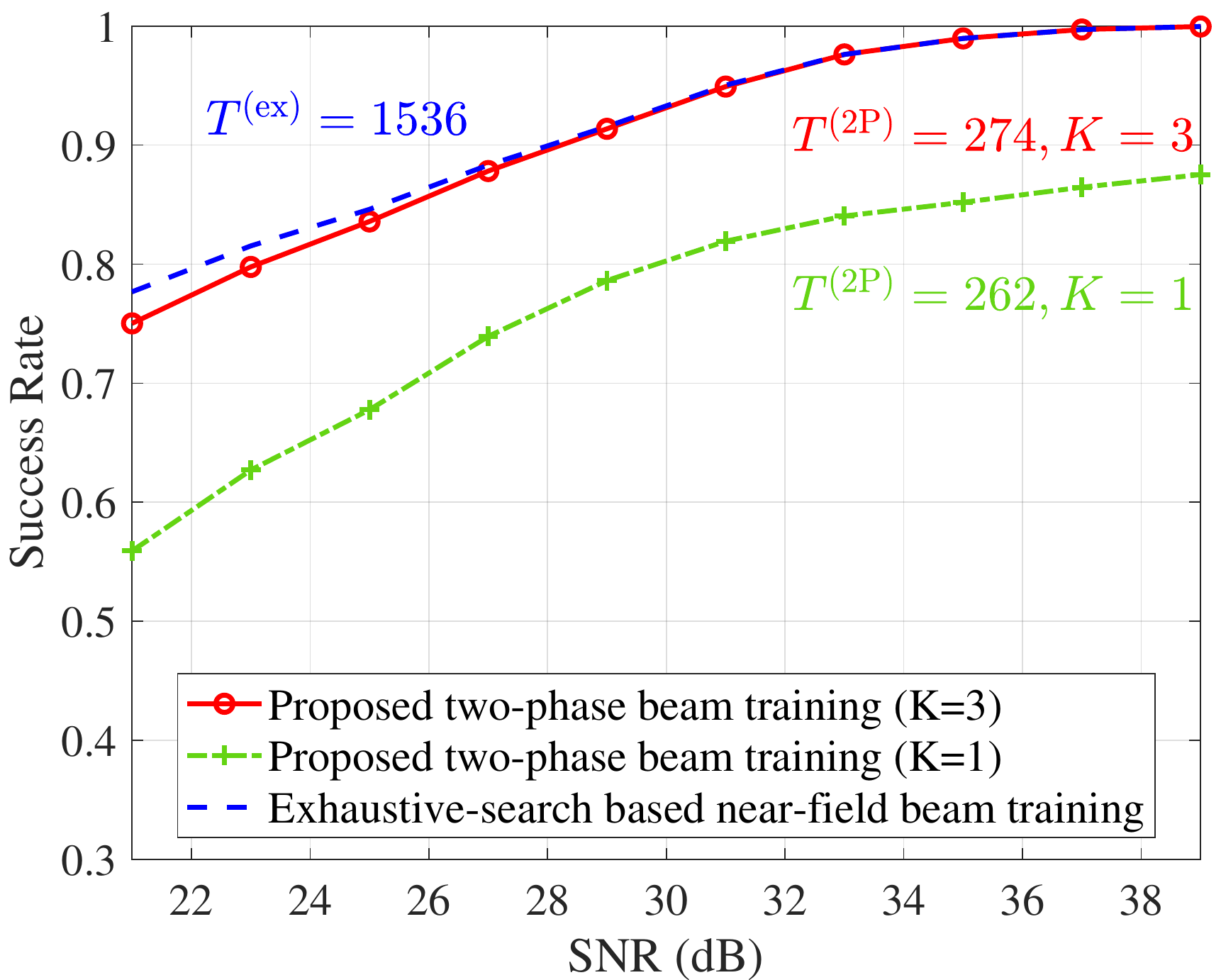}\label{SR_Vs_SNR}}
\subfigure[{Achievable rate versus SNR}]{\includegraphics[width=4.3cm]{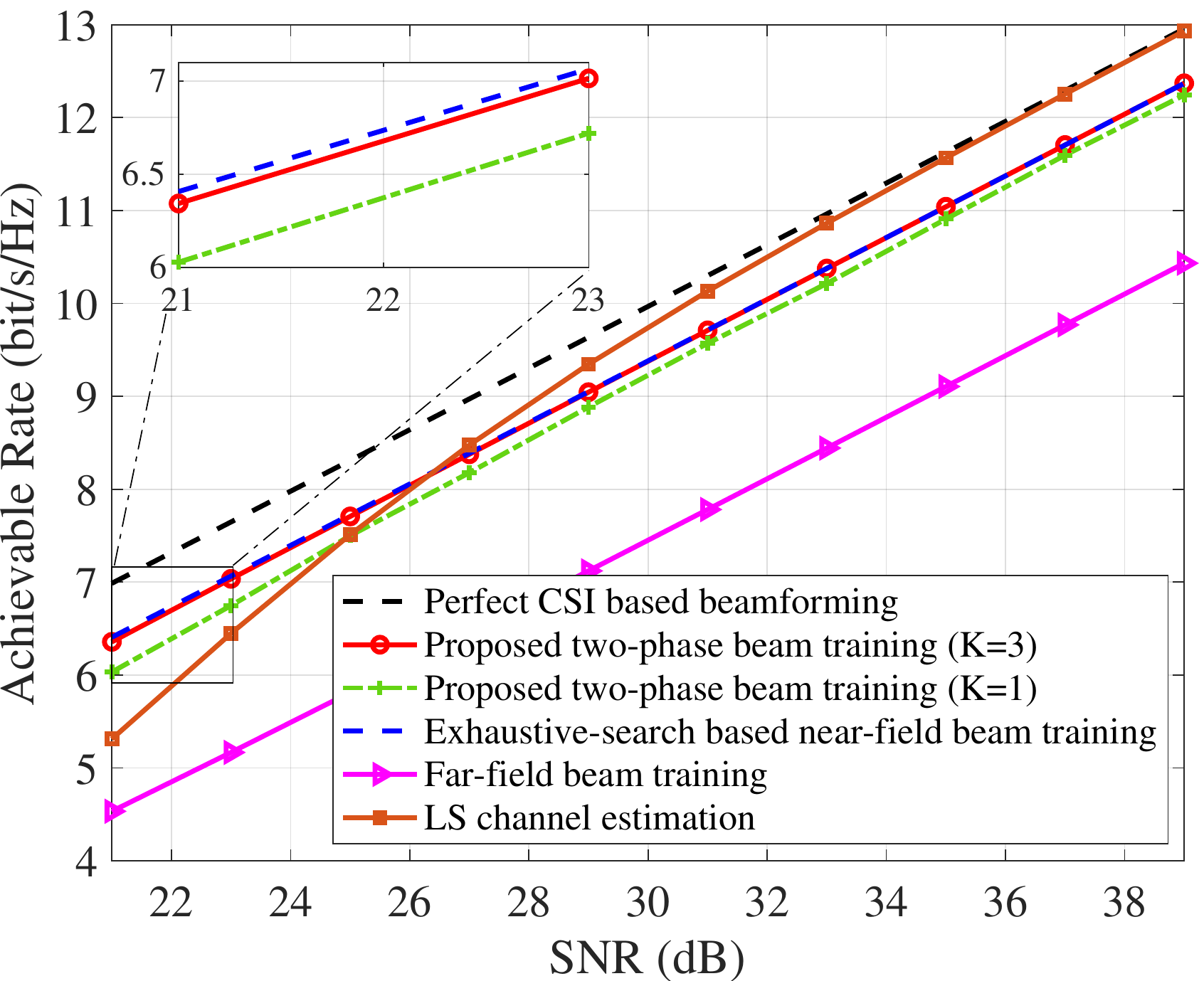}\label{AR_Vs_SNR}}
\subfigure[Success rate versus distance]{\includegraphics[width=4.3cm]{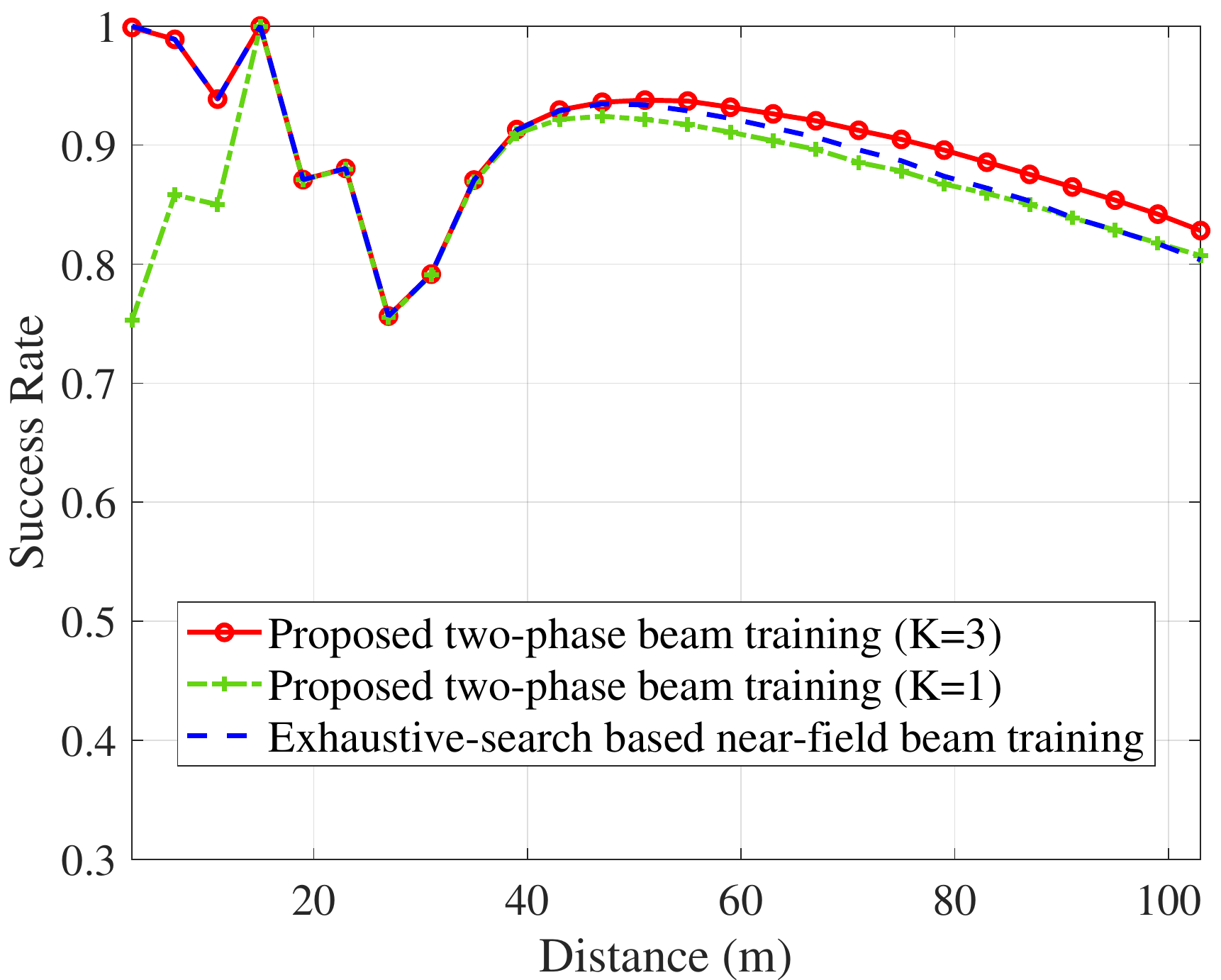}\label{SR_Vs_Dist}}
\subfigure[{Achievable rate versus distance}]{\includegraphics[width=4.3cm]{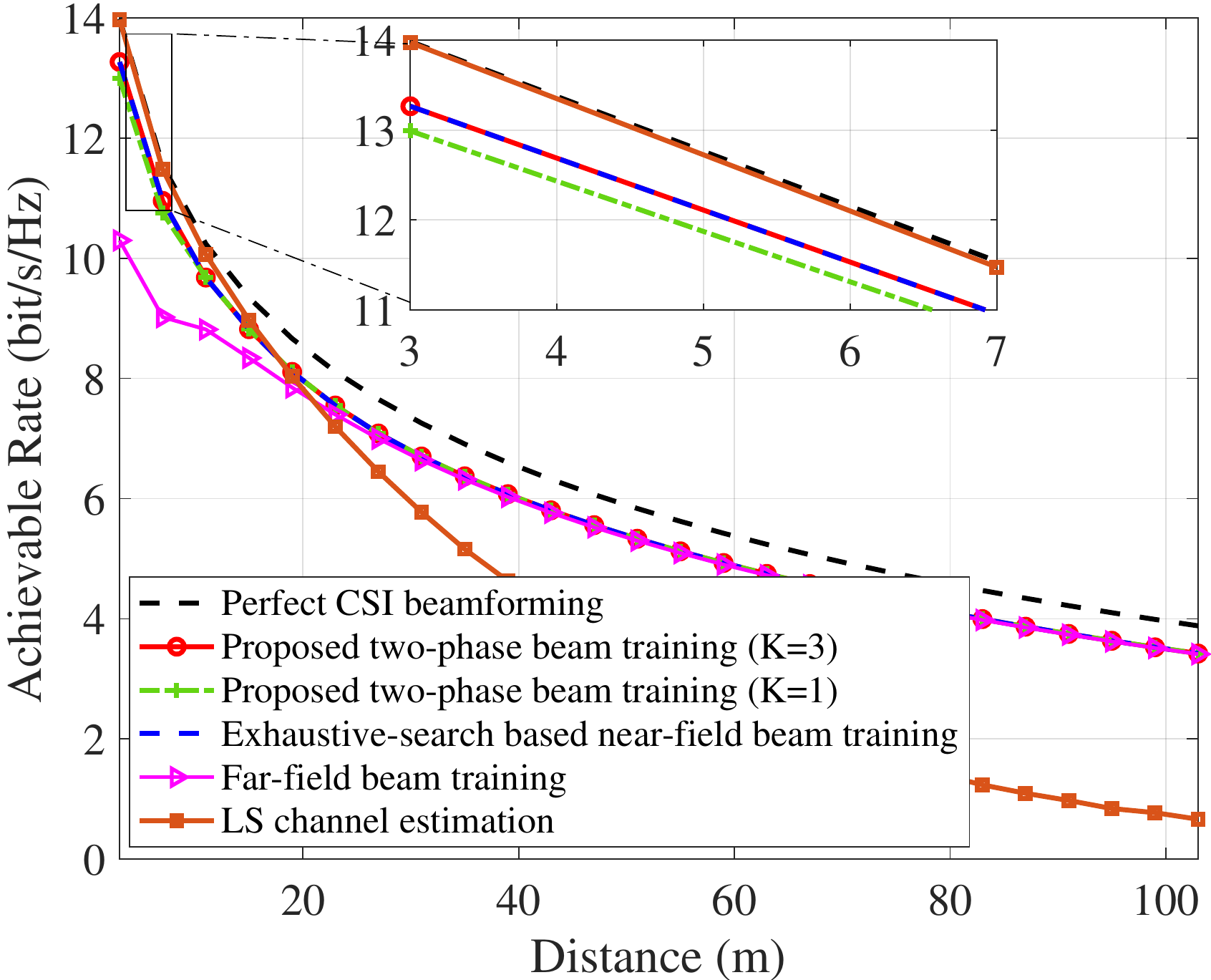}\label{AR_Vs_Dist}}
\vspace{-0.3cm}
\caption{Performance comparison of the proposed two-phase beam training with the exhaustive-search based near-field beam training and far-field beam training.}\label{simfig}
\vspace{-22pt}
\end{figure}

Figs.~\ref{SR_Vs_SNR} and \ref{AR_Vs_SNR} show the effects of the SNR on the success rate and achievable rate. The proposed two-phase beam training scheme is compared with the following three benchmark schemes: 1) \emph{perfect-CSI based beamforming}, which gives the rate performance upper bound, for which the transmit beamforming vector at the BS perfectly aligns with the user's channel; 2) \emph{exhaustive-search based near-field beam training}, for which the exhaustive search based near-field beam training is conducted based on the polar-domain near-field codebook, $\mathbf{F}$; 3) \emph{far-field beam training}, for which the conventional far-field codebook is adopted for beam training \cite{you2020fast}; {4) \emph{least-square (LS) downlink channel estimation method}, where the user estimates the channels with $N$ pilot symbols in the downlink.} Several interesting observations are made as follows. First, the proposed two-phase beam training scheme with $K=1$ greatly reduces the training overhead (i.e., $262$ versus $1536$), while it suffers a smaller success rate and slightly degraded achievable rate performance. This indicates that the proposed scheme still attains high-quality rate performance even though the BS beamforming may not perfectly align with the near-field channel. Second, by slightly increasing the number of candidate angles to $K=3$, the proposed beam training scheme attains a very close success rate as well as the achievable rate with the exhaustive-search based benchmark, while still achieving substantial training overhead reduction (i.e., $274$ versus $1536$). Moreover, it is observed that the proposed two-phase beam training method significantly outperforms the far-field beam training scheme under different SNRs. {In addition, it is worth noting that in the challenging low-SNR regime, our proposed beam-training method attains much better rate performance than the LS channel estimation method. This is because the beam training method can still form strong beams for effective beam determination, while the LS channel estimation method suffers a large estimation error due to low SNR.}

Finally, we show the effects of the user distance on the success rate and achievable rate in Figs.~\ref{SR_Vs_Dist} and \ref{AR_Vs_Dist}. The BS transmit power is set as $P=30$ dBm, the BS-user distance is ranging from $3$ m to $103$ m, and the user spatial angle is randomly distributed in $[-1,1]$. First, it is observed that with the increase of user distance, the success rate generally decreases while it fluctuates within a certain range (i.e., $15$--$40$ m) which can be explained as follows. With the non-uniform distance sampling method, when the user locates in the short-distance region, the success rate has a large value when the user is very close to the sampled distance points and then fluctuates within $15$--$40$ m due to the limited number of distance samples. On the other hand, in the long-distance region, the best effective distance estimation for the user is always the Rayleigh distance $D$, whose success rate is decreasing with the distance due to the larger path-loss. Second, the performance of the proposed scheme is very close to the exhaustive-search based near-field beam training at different distances and all the near- and far-field beam training schemes achieve the same rate performance when the distance is sufficiently large. Last, the achievable rates of all schemes monotonically decrease with the user distance due to the more severe path-loss.
\vspace{-14pt}
\section{Conclusion}
\vspace{-4pt}
In this letter, we proposed a novel two-phase near-field beam training method for the XL-array communication systems. It was shown that by sequentially performing beam sweeping in the angular and then distance domains, our proposed two-phase beam training method can significantly reduce the training overhead of the existing methods based on the two-dimensional exhaustive search, while achieving high-quality beamforming performance. Moreover, it is worth mentioning that the proposed two-phase beam training method is applicable to both the far- and near-field communication systems.

\bibliographystyle{IEEEtran}
\vspace{-10pt}
\bibliography{IEEEabrv,Ref}

\begin{thebibliography}{10}
\providecommand{\url}[1]{#1}
\csname url@samestyle\endcsname
\providecommand{\newblock}{\relax}
\providecommand{\bibinfo}[2]{#2}
\providecommand{\BIBentrySTDinterwordspacing}{\spaceskip=0pt\relax}
\providecommand{\BIBentryALTinterwordstretchfactor}{4}
\providecommand{\BIBentryALTinterwordspacing}{\spaceskip=\fontdimen2\font plus
\BIBentryALTinterwordstretchfactor\fontdimen3\font minus
  \fontdimen4\font\relax}
\providecommand{\BIBforeignlanguage}[2]{{%
\expandafter\ifx\csname l@#1\endcsname\relax
\typeout{** WARNING: IEEEtran.bst: No hyphenation pattern has been}%
\typeout{** loaded for the language `#1'. Using the pattern for}%
\typeout{** the default language instead.}%
\else
\language=\csname l@#1\endcsname
\fi
#2}}
\providecommand{\BIBdecl}{\relax}
\BIBdecl

\bibitem{cui2022near1}
M.~Cui, Z.~Wu, Y.~Lu, X.~Wei, and L.~Dai, ``Near-field communications for {6G}:
  Fundamentals, challenges, potentials, and future directions,'' \emph{arXiv
  preprint arXiv:2203.16318}, 2022.

\bibitem{9326394}
Q.~Wu, S.~Zhang, B.~Zheng, C.~You, and R.~Zhang, ``Intelligent reflecting
  surface-aided wireless communications: A tutorial,'' \emph{IEEE Trans.
  Commun.}, vol.~69, no.~5, pp. 3313--3351, 2021.

\bibitem{9617121}
H.~Lu and Y.~Zeng, ``Communicating with extremely large-scale array/surface:
  Unified modeling and performance analysis,'' \emph{IEEE Trans. Wireless
  Commun.}, vol.~21, no.~6, pp. 4039--4053, 2022.

\bibitem{zheng2022simultaneous}
B.~Zheng and R.~Zhang, ``Simultaneous transmit diversity and passive
  beamforming with large-scale intelligent reflecting surface,'' \emph{arXiv
  preprint arXiv:2202.04370}, 2022.

\bibitem{luo2022beam}
H.~Luo and F.~Gao, ``Beam squint assisted user localization in near-field
  communications systems,'' \emph{arXiv preprint arXiv:2205.11392}, 2022.

\bibitem{8949454}
Y.~Han, S.~Jin, C.-K. Wen, and X.~Ma, ``Channel estimation for extremely
  large-scale massive {MIMO} systems,'' \emph{IEEE Wireless Commun. Lett.},
  vol.~9, no.~5, pp. 633--637, 2020.

\bibitem{wei2021codebook}
X.~Wei, L.~Dai, Y.~Zhao, G.~Yu, and X.~Duan, ``Codebook design and beam
  training for extremely large-scale {RIS}: Far-field or near-field?''
  \emph{China Commun.}, vol.~19, no.~6, pp. 193--204, 2022.

\bibitem{cui2022channel}
M.~Cui and L.~Dai, ``Channel estimation for extremely large-scale {MIMO}:
  Far-field or near-field?'' \emph{IEEE Trans. Commun.}, vol.~70, no.~4, pp.
  2663--2677, 2022.

\bibitem{cui2022near}
M.~Cui, L.~Dai, Z.~Wang, S.~Zhou, and N.~Ge, ``Near-field rainbow: Wideband
  beam training for {XL-MIMO},'' \emph{arXiv preprint arXiv:2205.03543}, 2022.

\bibitem{you2020fast}
C.~You, B.~Zheng, and R.~Zhang, ``Fast beam training for {IRS}-assisted
  multiuser communications,'' \emph{IEEE Wireless Commun. Lett.}, vol.~9,
  no.~11, pp. 1845--1849, 2020.

\bibitem{7390101}
Z.~Xiao, T.~He, P.~Xia, and X.-G. Xia, ``Hierarchical codebook design for
  beamforming training in millimeter-wave communication,'' \emph{IEEE Trans.
  Wireless Commun.}, vol.~15, no.~5, pp. 3380--3392, 2016.

\bibitem{9738442}
H.~Zhang, N.~Shlezinger, F.~Guidi, D.~Dardari, M.~F. Imani, and Y.~C. Eldar,
  ``Beam focusing for near-field multi-user {MIMO} communications,'' \emph{IEEE
  Trans. Wireless Commun.}, pp. 1--1, 2022.

\end{thebibliography}

\end{document}